\documentclass[reprint, amsmath, amssymb, aps, prm, longbibliography,  superscriptaddress]{revtex4-2}

\usepackage{graphicx}
\usepackage{dcolumn}
\usepackage{bm}
\usepackage{amsmath}
\usepackage{xspace}
\newcommand{\RN}[1]{%
  \textup{\uppercase\expandafter{\romannumeral#1}}%
}

\DeclareUnicodeCharacter{2212}{-}
\begin{document}

\preprint{PRM}

\title{CNN-based TEM image denoising from first principles}

\author{Jinwoong Chae}
\affiliation{Department of Physics and HMC, Sejong University, Seoul, 05006, Republic of Korea}
\author{Sungwook Hong}
\affiliation{Department of Environment, Energy and Geoinfomatics, Sejong University, Seoul, 05006, Republic of Korea}
\author{Sungkyu Kim}
\affiliation{Department of Nanotechnology and Advanced Materials Engineering and HMC, Sejong University, Seoul, 05006, Republic of Korea}
\author{Sungroh Yoon}
\affiliation{Department of Electrical and Computer Engineering, Seoul National University, Seoul, 08826, Republic of Korea}
\author{Gunn Kim}
\altaffiliation{Corresponding author: gunnkim@sejong.ac.kr}
\affiliation{Department of Physics and HMC, Sejong University, Seoul, 05006, Republic of Korea}

\begin{abstract}
{Transmission electron microscope (TEM) images are often corrupted by noise, hindering their interpretation. To address this issue, we propose a deep learning-based approach using simulated images. Using density functional theory calculations with a set of pseudo-atomic orbital basis sets, we generate highly accurate ground truth images. We introduce four types of noise into these simulations to create realistic training datasets. Each type of noise is then used to train a separate convolutional neural network (CNN) model. Our results show that these CNNs are effective in reducing noise, even when applied to images with different noise levels than those used during training. However, we observe limitations in some cases, particularly in preserving the integrity of circular shapes and avoiding visible artifacts between image patches. To overcome these challenges, we propose alternative training strategies and future research directions. This study provides a valuable framework for training deep learning models for TEM image denoising.}
\end{abstract}
\maketitle

\section{Introduction}
Image and video noise poses a significant challenge in precision-dependent fields such as autonomous vehicles, robotic extraterrestrial exploration, and satellite imagery analysis, often obscuring critical information \cite{autonomous, Moseley_2021_CVPR, satellite_review}. In autonomous vehicles, for example, noise can interfere with accurate obstacle detection, while in satellite imagery it can prevent the identification of important geographic features \cite{satellite_review, satellite_denoise}. This problem is exacerbated by advances in nanotechnology, particularly in the analysis of two-dimensional images obtained by scanning tunneling microscopy (STM), transmission electron microscopy (TEM) and scanning electron microscopy (SEM) for the surface analysis of nanomaterials. In the contexts, noise is a significant barrier to the precise analysis required for technological progress.

Traditionally, mathematical noise reduction filters, including mean, Gaussian, and median filters, have been used to reduce image noise \cite{Mean_filter, Gaussian_filter, Median_filter}. These conventional methods typically operate by averaging or selecting the median of neighboring pixel values. However, they often suffer from significant limitations: inadvertently removing important image detail along with noise, degrading overall image quality, showing limited effectiveness against certain types of noise, and struggling with complex noise patterns \cite{IEEE.Tran.Img.Proc.24.9.1167(2002), IEEE.Tran.Patt.Anal.26.1.83(2004), CVPR.2833(2011)}. Moreover, optimizing these filters requires careful parameter tuning, a process that is both time-consuming and complex \cite{IEEE.Tran.Img.Proc.23.7.3114(2014), Fan2019}.

In recent years, deep learning(DL)-based noise reduction techniques have gained prominence due to their ability to preserve image integrity while effectively reducing noise \cite{N.N.131.251(2020), IEEE.Acc.9(2021)}. Among these, convolutional neural networks (CNNs) stand out for their ability to learn complex features of noise and image content through their layered architecture \cite{IEEE.Tran.Img.Proc.26.3142(2017)}. By training CNNs on large datasets of noisy and clean images, they can effectively remove noise while preserving important image details \cite{dnCNN01, dnCNN02, FFDNet}. This adaptability of CNNs allows noise reduction for various types of noise, which is effective for high-precision applications such as imaging the surface of nanomaterials \cite{Sci.Rep.11.5386(2021), IEEE.Tran.Comp.Img.8.585(2022), Phys.Rev.Mat.6.123802(2022), ICRAIC(2023)290}. 

CNN-based noise reduction methods have demonstrated impressive performance in denoising TEM images. However, their effectiveness is often hindered by the challenges associated with acquiring large training datasets. While experimental images provide real-world data, obtaining multiple measurements of the same sample is impractical, limiting the size and diversity of training sets. Therefore, researchers have explored hybrid approaches that combine CNNs with additional DL techniques \cite{ICRAIC(2023)290, JPCL.15.7.1985(2024)} or simulation-based methods \cite{IEEE.Tran.Comp.Img.8.585(2022), Phys.Rev.Mat.6.123802(2022), Ultra.219.113123(2020)}. Simulation-based methods have the advantage of generating a large amount of data with controlled noise levels. Some researchers have incorporated tight-binding calculations into their simulations to capture variations caused by changes in electronic structures that can affect image quality \cite{Phys.Rev.Mat.6.123802(2022)}. While tight-binding simulations provide the advantage of generating extensive data with precise noise control, their applicability is limited to a restricted set of materials. To overcome these limitations, density functional theory (DFT) can be used to provide a more versatile and adaptable approach.

In the present study, we first generate a large dataset using DFT calculations to train a highly accurate CNN for noise removal in two-dimensional nanomaterial images. We then introduce several types of noise into these images and apply the trained CNN to denoise them. CNNs have demonstrated remarkable success in image recognition and processing, and are expected to provide highly accurate and efficient noise removal capabilities \cite{Phys.Rev.Mat.6.123802(2022), Ultra.219.113123(2020), Appl.Mic.50.23(2020)}. Our approach differs from previous work by focusing on the specific noise characteristics of nanomaterial images, rather than optimizing the CNN architecture. This methodology enables accurate predictions with a relatively simple CNN model, providing both high accuracy and computational efficiency.

\section{Data and Methods}
\subsection{First-principles Calculations}

We employed DFT calculations, implemented using the open-source OpenMX code, to obtain the electronic charge density of a graphene monolayer \cite{Phys.Rev.B.67.155108(2003)}. A pseudoatomic orbital (PAO) basis set was used to describe the electronic charge density, while norm-conserving pseudopotentials were employed to represent the electron-ion interactions \cite{Phys.Rev.Lett.43.1494(1979), Phys.Rev.Lett.48.1425(1982)}. The exchange-correlation energy was calculated using the Perdew-Burke-Ernzerhof (PBE) generalized gradient approximation (GGA) functional \cite{Phys.Rev.Lett.77.3865(1996)}. A kinetic energy cutoff of 300 Ry and a convergence tolerance of \(10^{-8}\) Hartree were chosen. Given the structural similarity between graphene and transition metal dichalcogenides (TMDCs), we tried constructing a training dataset based on the graphene structure. For the electronic structure calculations of graphene, a \(\Gamma\)-centered \(18 \times 18 \times 1\) k-point grid was used. Finally, the model structures were visualized, and two-dimensional charge density maps were plotted using the VESTA, Gnuplot, and Matplotlib packages.

\subsection{Generation of Disordered Atomic Configurations}
Real-world TEM samples often deviate from idealized crystal structures, exhibiting surface imperfections. To ensure that our simulated ground truth (GT) images accurately represent these variations, we randomly arranged atomic configurations rather than relying on perfectly periodic structures. By allowing atoms to deviate from their ideal positions within a certain range, we introduced structural variability similar to that observed in experimental TEM images. We set a maximum displacement of \(\pm 20\%\) of the bond length in the in-plane coordinates (x and y). For a graphene monolayer, this translated to a maximum positional shift of 0.284 {\AA} for each atom in both the x and y directions. This approach enhances the fidelity of our simulated images to real TEM data, thereby improving the robustness and generalizability of our noise reduction method.

To modulate the displacement of the atoms, we introduced coefficient values (\(C_G\)) using a Gaussian distribution (\(val_G\)) with a mean of 0.0 and a standard deviation of 1.0. When generating disordered structures, the positions of atoms were shifted by changing the \(C_G\) at each atom in the lattice. The \(C_G\) generated from a \(val_G\) allows the atoms to move beyond their maximum. Therefore, we adjusted the \(C_G\) to stay in the range of \(\pm 1\), as follows:
\begin{eqnarray}
C_G = \begin{cases}
            val_{G}/5, & \text{if} ~ 0 \leq |val_{G}| \leq 1 \sigma \\
            val_{G}/4, & \text{if} ~ 1 \sigma < |val_{G}| \leq 2 \sigma \\
            val_{G}/3, & \text{if} ~ 2 \sigma < |val_{G}| \leq 3 \sigma \\
            -1, & \text{if} ~ val_{G} < -3 \sigma \\
            +1, & \text{if} ~ 3 \sigma < val_{G} \\
      \end{cases}
\end{eqnarray}
where \(\sigma\) is the standard deviation. The \(C_G\) multiplies with the maximum shifted distance (0.284 {\AA}) to determine the new atomic configurations. 

Thus, the alternative positions of the atoms are given by the following equation:
\begin{eqnarray}
    P_{disorder}(x, y) = 0.284 \cdot C_{G} + P_{order}(x, y),
\end{eqnarray}
where \(P_{disorder}\) is the alternative position of the atom and \(P_{order}\) is the position of the atom in the ordered structure. A significant majority (68.2\%) of atoms in the lattice exhibit positional shifts of less than 4\% of the bond length. Approximately 27.2\% of atoms experience moderate displacements between 5\% and 10\% of the bond length. The remaining 4.6\% of atoms undergo more substantial modifications, with positional shifts ranging from 13.3\% to 20\% of the bond length. One of the randomly distributed atomic configurations is presented in Fig.~\ref{fig:fig_schematics}(a). 

Next, we obtained electron densities of the disordered structures from the DFT calculation and created charge density maps. The charge density maps indicate the electron density 3 {\AA} above the surface in the z-direction. Interestingly, even at the same tip position, the bright spots corresponding to the locations of the atoms in the charge density map have different brightness and sizes due to the changes in bond lengths. The red rectangles in Fig.~\ref{fig:fig_schematics}(a) highlight the areas where differences in brightness are observed.

\subsection{Large-scale Images from Randomly Arranged Charge Density Maps}
Experimental TEM images often focus on structures at the nanometer scale. First-principles calculations for electron charge densities in such large-scale systems are time-consuming and computationally intensive. To address this limitation, we exploit the efficiency of the CNN method to significantly reduce computational costs. 
A key advantage of CNNs is their ability to process images directly, without requiring the full electron charge density of large-scale systems. Using this, we introduce a novel approach: instead of calculating the charge density for entire systems, we combine charge density maps of smaller, disordered structures to generate representative images. This method effectively simulates large-scale systems while maintaining computational efficiency. 
A schematic of the lateral combining process to obtain images of large-scale systems is shown in Fig.~\ref{fig:fig_schematics}(b). Combining charge density maps from disordered structures can increase the prevalence of extended charge density maps. While our schematic shows a 3 \(\times\) 3 configuration, the image size can be adjusted to 2 \(\times\) 2, 4 \(\times\) 4, or other variations. This approach remarkably reduces the computational costs associated with generating diverse images of large-scale systems.

\subsection{Shifting Image for Edge Description}
To enhance training efficiency, we adopted an image-splitting method. The original image, sized at $512 \times 512$ pixels, was divided into $8 \times 8$ patches, each measuring $64 \times 64$ pixels. A challenge arises when dealing with bright spots located at the edges of the split images. These bright spots appear as truncated circles due to the splitting, and if only a few such shapes are present, the model may produce predictions that deviate from the original image. To address this issue and increase the diversity of truncated circles in the training set, we generated new images by shifting the original image by a few pixels along both the x and y axes. The schematic for generating the shifted images is shown in Fig.~\ref{fig:fig_schematics}(c). The shift parameter \(d\) specifies the amount by which the image is shifted along the x and y directions, effectively dividing the image into four regions labeled i, ii, iii, and iv. By rearranging these regions under the assumption of a repeated structure, the shifted image is constructed.
Furthermore, an image equivalent in size to the training set images is shown in Figure~\ref{fig:fig_schematics}(c). A magenta rectangle represents one of the truncated circles at the edge of the image. In this study, the parameter \(d\) was set to \(8\), \(10\), and \(12\) pixels to generate three shifted images from each original image.

\subsection{Generation of Noisy Images}
\begin{figure*}[t!]
\centerline{\includegraphics[width=.9\textwidth]{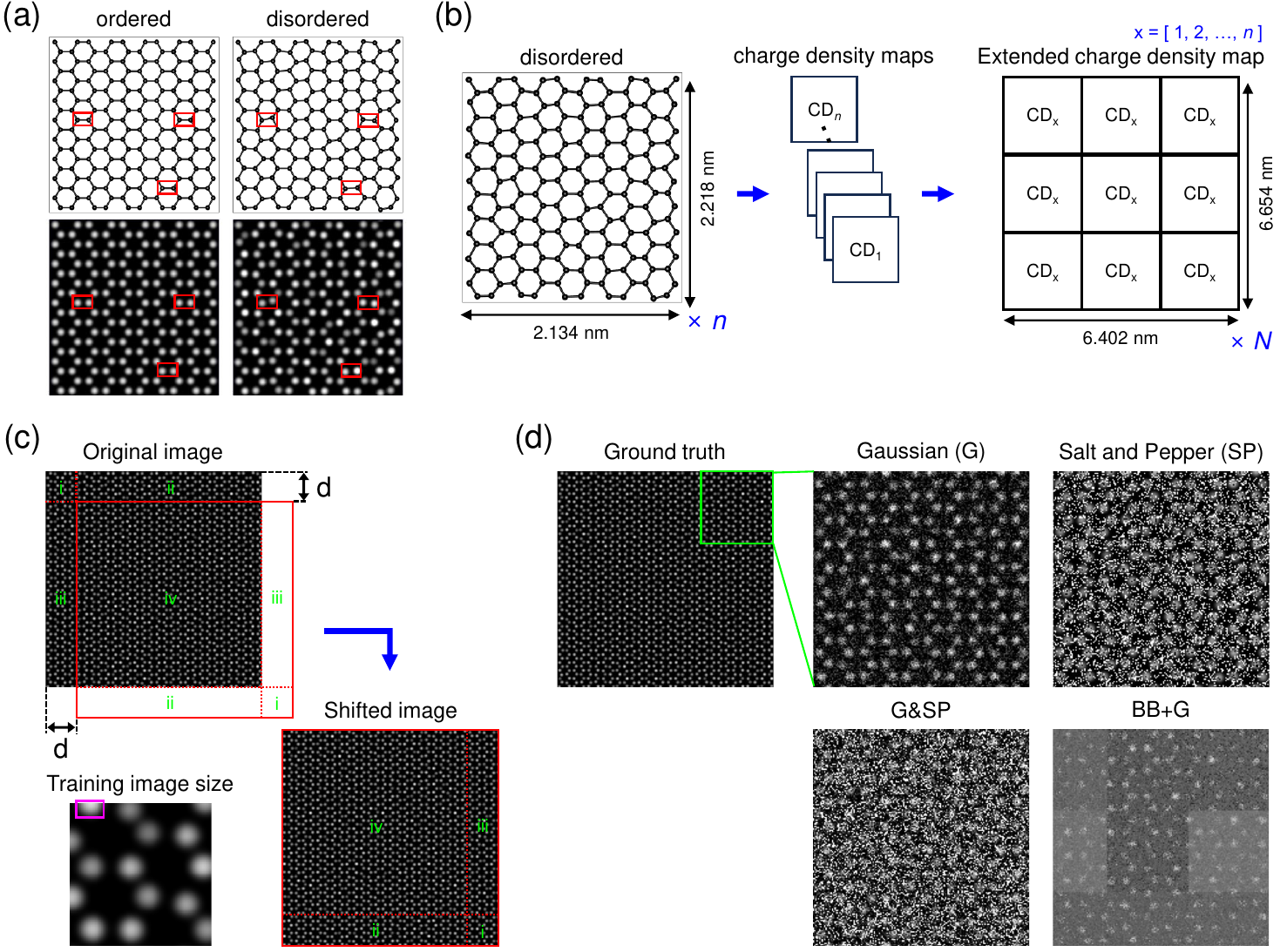}}
\caption{(a) Atomic configurations and charge density maps. The red rectangles indicate the regions that differ from each other. 
(b) Schematic of the process for obtaining the lateral combined charge density map. Here, \(n\) represents the number of disordered structures, and \(N\) represents the number of lateral combined charge density map images. (c) Scheme for generating shifted images. The magenta rectangle indicates the truncated circle at the edge of the image. (d) Enlarged images illustrating the generated noise in detail. The ground truth image shows one of the lateral combined charge density maps, with the green square indicating the magnified region. The right panel image, labeled BB\(+\)G, shows the image with applied background brightness and Gaussian noise.}
\label{fig:fig_schematics}
\end{figure*}

To determine the appropriate noise for generating training sets for TEM-denoising CNN model training, we considered three types of noise: Gaussian (G) noise, salt-and-pepper (SP) noise, and a combination of both (G$\&$SP). The effects of these noise types are illustrated in Figure~\ref{fig:fig_schematics}(d). These noises are applied to the simulated ground truth (GT) images to produce noisy images. Prior to noise generation, all pixel values are normalized from 0 to 1 by dividing by 255 -- the maximum channel value in gray-scale images. Additionally, we introduce a coefficient to control the magnitude of the noise signal, denoted as the noise factor \(C_{\text{noise}}\).

For Gaussian noise, \(C_{\text{noise}}\) was multiplied by a random value from a normal distribution (\(val_{RN}\)) to generate the noise. The pixel values were modified according to the following equation:
\begin{eqnarray}
 I_{\text{noisy}} = I_{GT} + C_{\text{noise}} \cdot val_{RN},
\end{eqnarray}
where \(I_{\text{noisy}}\) is the pixel value of the noisy image and \(I_{GT}\) is the ground truth image. Because \(I_{\text{noisy}}\) may fall outside the normalized pixel value range of 0 to 1, we applied a truncation process to set any pixel values below 0 to 0, and any values above 1 to 1. To generate SP noise, we used a probability factor \(p\) that determines the likelihood of a pixel being altered. For each pixel, a random number from a uniform distribution between 0 and 1 was generated. If this random number is less than \(p\), the pixel's value is replaced with either 0 (black) or 1 (white). Specifically, if the original pixel value is below 0.5, it is set to 1 (white); if it is 0.5 or above, it is set to 0 (black). For the combination of Gaussian and SP noise (G\&SP), the image was first corrupted with Gaussian noise and then with SP noise.

\subsection{Background Brightness}
The brightness of the spots varies depending on the background brightness in each region; bright spots appear less intense in dark background areas and more pronounced in bright background areas. It is essential to consider these background brightness variations when generating noisy images for model training. Since we divide the corrupted images into 8 $\times$ 8 patches for efficient model training, we apply different background brightness levels to each patch individually.

To account for variations in background brightness (BB), we implemented a two-stage process: (1) BB correction and (2) modification of bright spot brightness. In the first step, we considered three different BB correction types—dark, slightly bright, and high-brightness—introduced using uniform random distributions. We employed two uniform random variables, \(u^{a}\) and \(u^{b}\). The variable \(u^{a}\) determines the number of patches assigned to each BB correction type, while \(u^{b}\) specifies the brightness benchmark for these patches. The range of \(u^{b}\) is influenced by \(u^{a}\) and a hyperparameter \(w_{1}\). In this study, the probabilities for both the dark and high-brightness BB correction types were set to 15\%, and the hyperparameter \(w_{1}\) was set to 0.2. We defined the range of \(u^{b}\) as follows:
\begin{eqnarray}
u^b = \begin{cases}
            [0.4 \cdot w_1:0.6 \cdot w_1], & \text{if} ~ 0 \leq u^a < 0.15 \\
            [0.8 \cdot w_1:1.2 \cdot w_1], & \text{if} ~ 0.85 \leq u^a \leq 1 \\
            [0.6 \cdot w_1:0.9 \cdot w_1], & \text{otherwise} \\
      \end{cases}
\end{eqnarray}
where the first case indicates the dark BB effect, while the second case demonstrates the high-brightness BB effect.

We modified the brightness of the bright spots by applying a Gaussian distribution with a mean of \(u^{b}\) and a standard deviation of the hyperparameter \(w_{2}\). The value obtained from this Gaussian distribution, denoted as \(val_{BB}\), determines a threshold for pixel intensities. If a pixel's value is lower than \(val_{BB}\), it is replaced with \(val_{BB}\), effectively brightening darker areas to create a blurred background, as shown in Figure~\ref{fig:fig_schematics}(d). Pixels with values higher than \(val_{BB}\) are considered to be part of or near the bright spots. To further adjust the pixel values to resemble the blurred background brightness, we introduced a smoothness coefficient \(C_{BB}\). This coefficient is a variable parameter that depends on the pixel's intensity and the value of \(u^{a}\). The modified pixel values after applying the background brightness effect, denoted as \(I_{BB}\), are calculated as follows:

\begin{eqnarray}
I_{BB} = \begin{cases}
            val_{BB}, & \text{if} ~ I_{GT} \leq val_{BB} \\
            I_{GT} - C_{BB} \cdot val_{BB}, & \text{otherwise} \\
      \end{cases}
\end{eqnarray}
where \(I_{GT}\) is the pixel value of the ground truth image. We also generated the corrupted images by sequentially applying background brightness and Gaussian noise. The resulting noisy image is shown in the right panel of Fig.~\ref{fig:fig_schematics}(d) labeled BB\(+\)G.

\subsection{Convolution Neural Networks (CNN)}
We implemented the CNN model in Python using the Keras~\cite{KERAS} library with TensorFlow~\cite{tensorflow} as the backend. The architecture is based on the U-Net network~\cite{U-Net} and is influenced by previous research~\cite{Phys.Rev.Mat.6.123802(2022)}. The model consists of a series of building blocks: two downsampling blocks, a central bottleneck block, and two upsampling blocks. The downsampling blocks use the Keras \texttt{MaxPooling} operation to halve the array sizes, while the upsampling blocks employ the \texttt{UpSampling} operation to double the array sizes. Each block consists of two convolutional layers with rectified linear unit (ReLU) activation functions, each followed by batch normalization. We used the ReLU activation function in the final layer of our architecture to ensure non-negative output values. The CNN model is optimized using the Adam algorithm~\cite{ADAM} in Keras and trained with the mean absolute error as the loss function. The learning rate was set to 0.001, and the number of training epochs was 150. We used 20\% of the dataset for validation and the remaining 80\% for training. We generated approximately \(1.1 \times 10^{4}\) images, using \(9 \times 10^{3}\) of them for model training.

\begin{figure*}[t]
\centerline{\includegraphics[width=.9\textwidth]{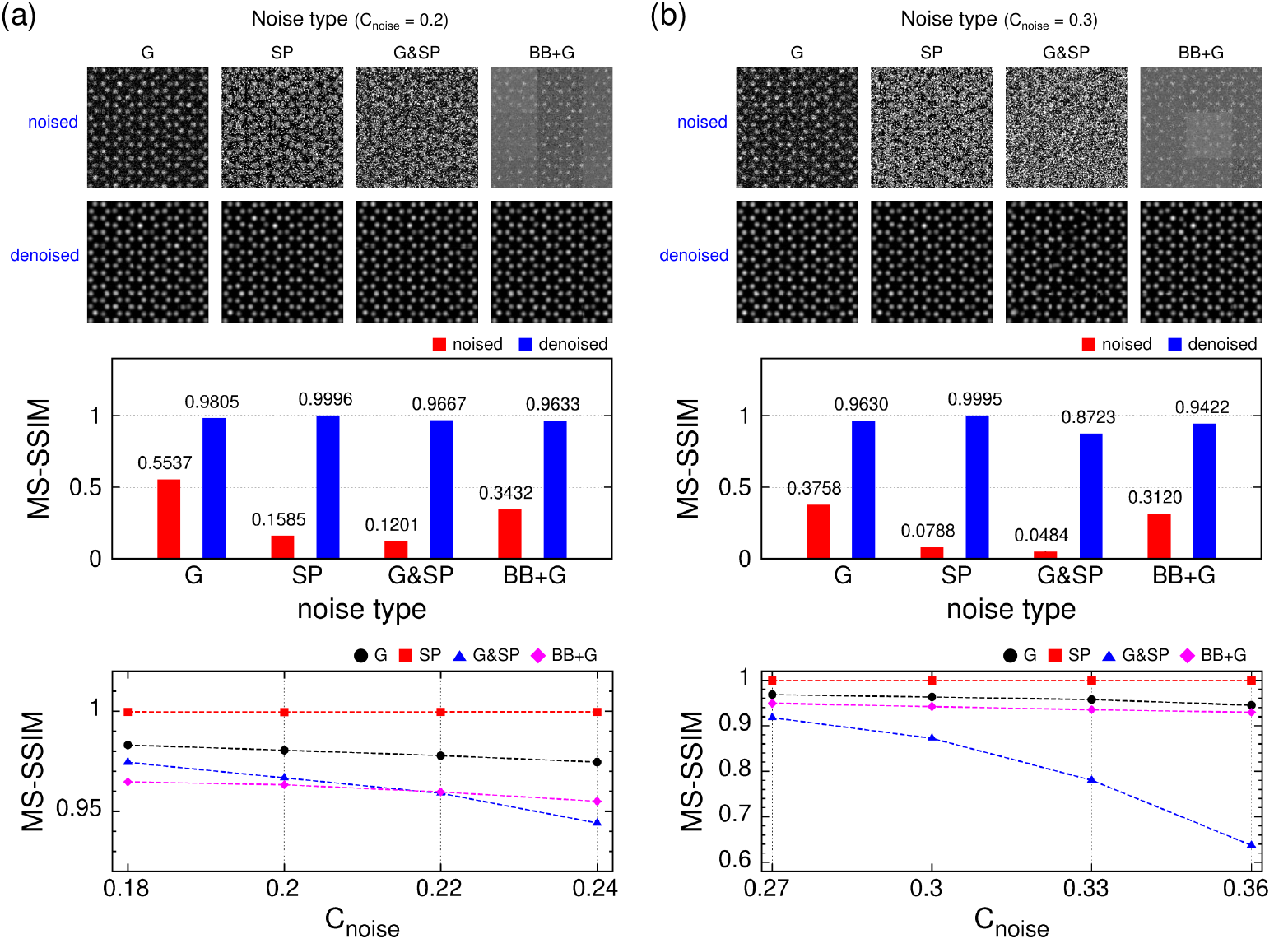}}
\caption{(a) Enlarged sections of the corrupted and predicted images with \(C_{\text{noise}} = 0.2\) and the corresponding MS-SSIM chart. 
(b) Enlarged sections of the corrupted and predicted images with \(C_{\text{noise}} = 0.3\) and the corresponding MS-SSIM chart. 
The lowest panel shows MS-SSIM values as a function of \(C_{\text{noise}}\) for both the interpolation and extrapolation regions.}
\label{fig:fig_model_test}
\end{figure*}

\section{Results and Discussion}
\subsection{Preparation for Training DL Model}
As previously mentioned, we considered a graphene monolayer sheet with a honeycomb lattice structure in this study, since scanning transmission electron microscopy (STEM) images of \(\mathrm{MoS}_2\) reveal hexagonal patterns. We prepared a total of six disordered structures and obtained charge density maps for each. One structure was reserved for the test set, while the remaining structures were used to generate the training sets.

After obtaining the charge density maps of the disordered structures, we generated laterally combined images of different sizes to train the CNN model: five images of \(3 \times 3\), three images of \(4 \times 4\), and three images of \(2 \times 2\). We focused on identifying the appropriate noise type to employ in the training sets for the denoising model. Therefore, we considered four models, each trained using one of the noise types presented in Figure~\ref{fig:fig_schematics}(d). Various noise factors (\(C_{\text{noise}}\)) were used to generate noisy images for the training sets, aiming to enhance the model's performance. We set the \(C_{\text{noise}}\) values to 0.1, 0.15, 0.2, and 0.25. We evaluated the performance of the models by measuring the multi-scale structural similarity index (MS-SSIM)~\cite{SSIM, IEEE.Trans.Comp.Imag.3.1.47} of the predicted images.

\subsection{Prediction Results: Interpolation and Extrapolation}
This study evaluates our CNN models prediction results using the SSIM in both interpolation and extrapolation regions. Figure~\ref{fig:fig_model_test} presents the MS-SSIM plots for each model's interpolation and extrapolation regions, including the corrupted and predicted images, displayed as enlarged sections of the full-sized images. In addition, Figure~\ref{fig:fig_model_test} presents the corrupted and predicted images, which are enlarged portions of the full-size image. As shown in Figure~\ref{fig:fig_model_test}(a), the noise factor \(C_{\text{noise}} = 0.2\) for the test images, which is one of the values used in the training sets. In this case, all the MS-SSIM values for the predicted images are above 0.96, an expected result due to the matching \(C_{\text{noise}}\) of the test images.

To further evaluate our model, we used different \(C_{\text{noise}}\) values than those used to generate the training sets. We selected \(C_{\text{noise}} = 0.18\), \(0.22\), and \(0.24\) for the interpolation region. Figure~\ref{fig:fig_model_test}(a) shows the MS-SSIM values for various \(C_{\text{noise}}\) in the interpolation region. As the \(C_{\text{noise}}\) value increases, the MS-SSIM value for the G\&SP model slightly decreases, while the MS-SSIM values for the G, SP, and BB\(+\)G models remain nearly constant. Interestingly, in the interpolation region, all the predicted images from our models have MS-SSIM values above 0.94. These results indicate that each model is well-trained for its respective type of noise.

To evaluate the models under extrapolation conditions, we generated test images with a noise factor \(C_{\text{noise}} = 0.3\), a value not used during training. The test images and the corresponding predictions by the models are shown in Figure~\ref{fig:fig_model_test}(b). Interestingly, all the MS-SSIM values for the predicted images are around 0.9. The model trained with combined Gaussian and salt-and-pepper noise (G\&SP) shows the lowest MS-SSIM value compared to the other models, while the highest MS-SSIM was observed in the SP noise case. 

Furthermore, we evaluated the extrapolation ability of our models using corrupted images generated with different \(C_{\text{noise}}\) values of 0.27, 0.33, and 0.36. All models, except the G\&SP model, performed well in the extrapolation region. The MS-SSIM values of these models (G, SP, and BB\(+\)G) remained higher than 0.9 in all cases. As \(C_{\text{noise}}\) increases, the MS-SSIM values for the predictions by the G\&SP model decreased to $\approx$ 0.6. The results indicate the potential of our models to eliminate noise in experimental images.

Since the test images were generated using the same noise type as the training set, it is challenging to fully assess the model's performance based only on the MS-SSIM values of the predicted images. One of our primary focuses in this study is finding the appropriate noise type for generating training sets. Thus, our model testing scheme addresses the issue of MS-SSIM values not fully reflecting model performance. Another goal is to reduce the number of training images required. We used approximately \(9 \times 10^{3}\) images to train the models, and the predicted images are comparable to the original images. The full-size corrupted images used in the test sets with \(C_{\text{noise}} = 0.3\) are shown in Figure~\ref{fig:fig_S1}.

\begin{figure*}[t]
\centerline{\includegraphics[width=.9\textwidth]{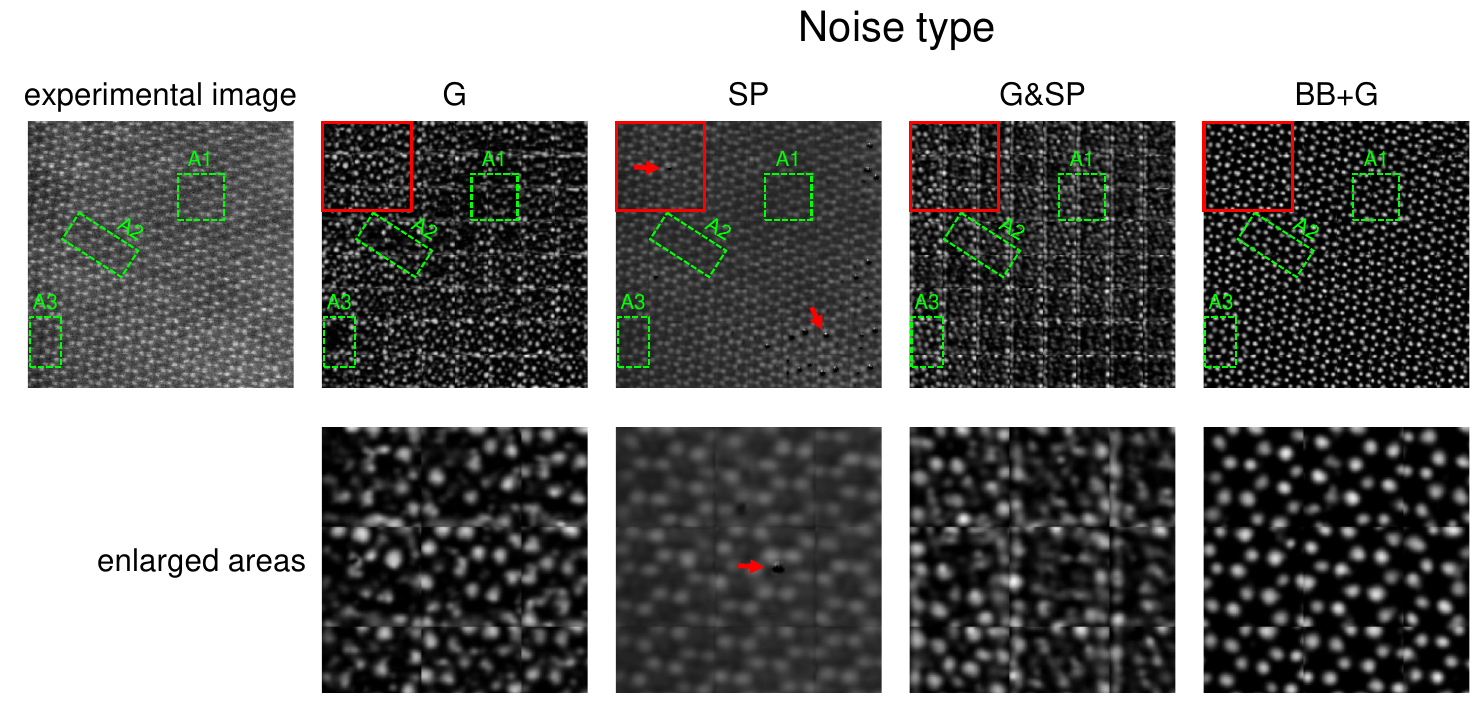}}
\caption{Predicted images from experimental data using the models. The red squares show enlarged areas in full-size images. The regions labeled A1, A2, and A3 indicate low-brightness areas in the experimental TEM image. The red arrows point to the bubble-shaped defects generated when predicting extrapolated regions.}
\label{fig:fig_exp_denoise}
\end{figure*}

\subsection{Denoising Experimental Image}
Despite training our four models exclusively on graphene monolayer sheets, the hexagonal patterns in graphene and \(\mathrm{MoS}_2\) TEM images show similarities. Taking advantage of this, we applied our models to denoise an experimental \(\mathrm{MoS}_2\) monolayer image. The predicted results are shown in Figure~\ref{fig:fig_exp_denoise}, showing both full-size and enlarged portions of the denoised images.

The predicted images obtained from the SP and BB\(+\)G models accurately depict the hexagonal patterns, whereas those from the G and G\&SP models reveal messy patterns due to the spreading of bright spots. Although the G model shows extrapolation ability, the messy patterns in its predictions emphasize the limitations of our model testing scheme. Interestingly, the predicted images from the SP and BB\(+\)G models exhibit different features. In the SP model's denoised images, the bright spots are predicted with uniform sizes and shapes, whereas the BB\(+\)G model predicts bright spots with varying sizes and shapes. In particular, some of the bright spots in the image predicted by the BB\(+\)G model have a circular shape with tails. Moreover, the BB\(+\)G model effectively separates the bright spots from the background, whereas the image predicted by the SP model appears blurry. The SP model's prediction is less clear than the experimental TEM image and exhibits bubble-shaped defects at certain points, as indicated by the red arrows in Figure~\ref{fig:fig_exp_denoise}.

The TEM images exhibit subtle variations between bright and dark areas, and the brightness variations contain structural information about the material. Therefore, we can validate the CNN model's performance by examining the areas with brightness differentiation in the predicted images. The predicted low-brightness areas confirm our models' ability to extrapolate. We denote the dark areas in Figure~\ref{fig:fig_exp_denoise} using green dashed rectangles (A1, A2, and A3). The G and G\&SP models show poor performance in predicting the brightness of different areas. The G model fails to predict the bright spots, and the G\&SP model generates bright spots in incorrect locations. Messy patterns appear in A1 with the G model, while blank areas appear in A2 and A3. The G\&SP model creates bright spots in the centers of the hexagonal patterns, as shown in all the rectangles. The SP model can distinguish the areas according to their brightness, although it predicts a blurry image. The BB\(+\)G model accurately describes the hexagonal patterns, even within the rectangles. However, it fails to reproduce the brightness differences between the inside and outside of the rectangles. During the generation of training sets for the BB\(+\)G model, our focus was on representing brightness differentiation by area. As a result, the BB\(+\)G model's predictions closely resemble the GT images used in the training sets, rather than capturing the brightness differentiation between regions.

\section{Conclusion}

In this study, we proposed a novel CNN-based approach for effectively removing noise from TEM images, even with limited training data. Our models demonstrated robust performance across various noise types, including Gaussian, salt-and-pepper, and combinations thereof. By training specialized models for each noise type, we achieved significant improvements in image quality. While we found that the optimal noise model varied depending on specific imaging conditions, our results show the versatility of deep learning for addressing noise in TEM imaging. We confirmed the model's denoise ability in the interpolation and extrapolation regions by measuring the multi-scale structural similarity index (MS-SSIM) values. 

Our work differs from previous studies that have relied only on tight-binding calculations for simulation-based noise reduction. While tight-binding simulations offer precise noise control, their applicability is limited to a restricted set of materials. By employing DFT, we provide a more versatile framework for capturing electronic structure variations across a broader spectrum of materials. This versatility facilitated a more comprehensive exploration of how structural changes influence image quality, leading to a more effective noise reduction solution.

To address the limitations of our current approach, our future research will focus on several key areas. First, we plan to generate training sets with sliced patches that incorporate various interatomic distances at the pixel scale. This adaptation will ensure that the predicted images accurately represent the spatial distribution of bright spots, thus eliminating the tailed circular artifacts observed in our current results. Second, we will explore alternative patch generation methods, such as random patch selection, to mitigate the horizontal and vertical lines introduced by the sequential slicing scheme. This will help reduce artifacts in the reconstructed images and improve the overall quality of the denoised images. Finally, we will investigate advanced deep learning techniques, such as generative adversarial networks and active learning, to further improve denoising performance. By addressing these challenges, we aim to develop a more comprehensive and versatile noise reduction framework that can be applied to a wider range of microscopy applications.

\section{Acknowledgments}
J.C. and G.K. acknowledge the financial support from the Basic Science Research Program through the National Research Foundation of Korea (NRF), funded by the Government of Korea (Grant No. NRF-2020R1A6A1A03043435).

\appendix
\section{}
\begin{figure*}
\centerline{\includegraphics[width=.7\textwidth]{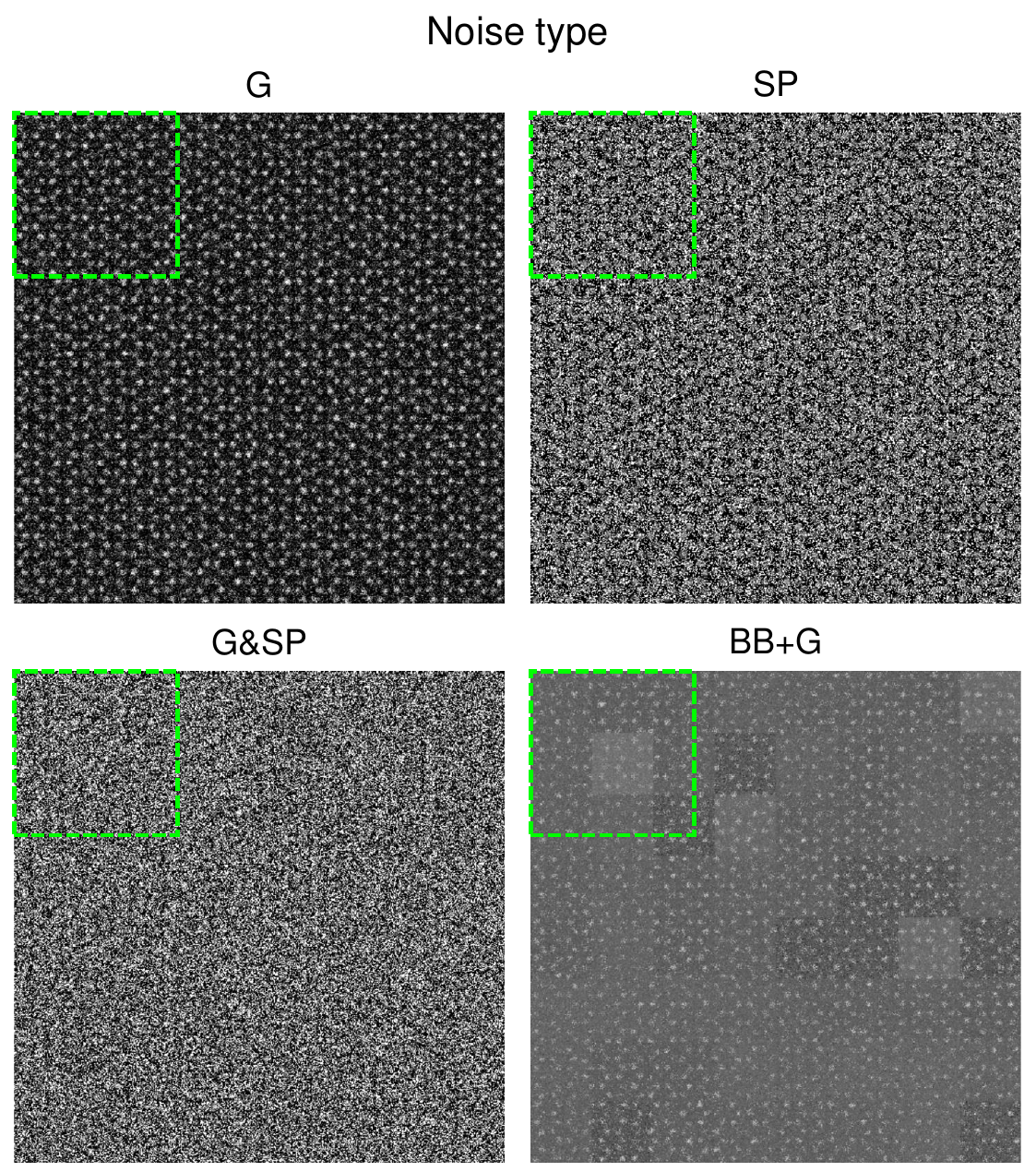}}
\caption{Full-size corrupted images used in the test sets, with the noise coefficient $C_{\text{noise}}$ set to 0.3. The green dashed squares indicate enlarged sections, which are presented in Figure~\ref{fig:fig_model_test}(b).}
\label{fig:fig_S1}
\end{figure*}

In Figure~\ref{fig:fig_S1}, we present the full-size corrupted images used as test sets, where the noise coefficient \( C_{\text{noise}} \) is set to 0.3. The images are utilized to evaluate the extrapolation regions of the CNN models. The labels G, SP, G\&SP, and BB\(+\)G indicate the types of noise applied: Gaussian (G), salt-and-pepper (SP), Gaussian and salt-and-pepper combined (G\&SP), and background brightness with Gaussian noise (BB\(+\)G), respectively. The full-size images have dimensions of \( 6.402 ~\times ~6.654 \) nm\(^2\) (width\(~\times\)~height).

\section{}
\begin{figure*}
\centerline{\includegraphics[width=.7\textwidth]{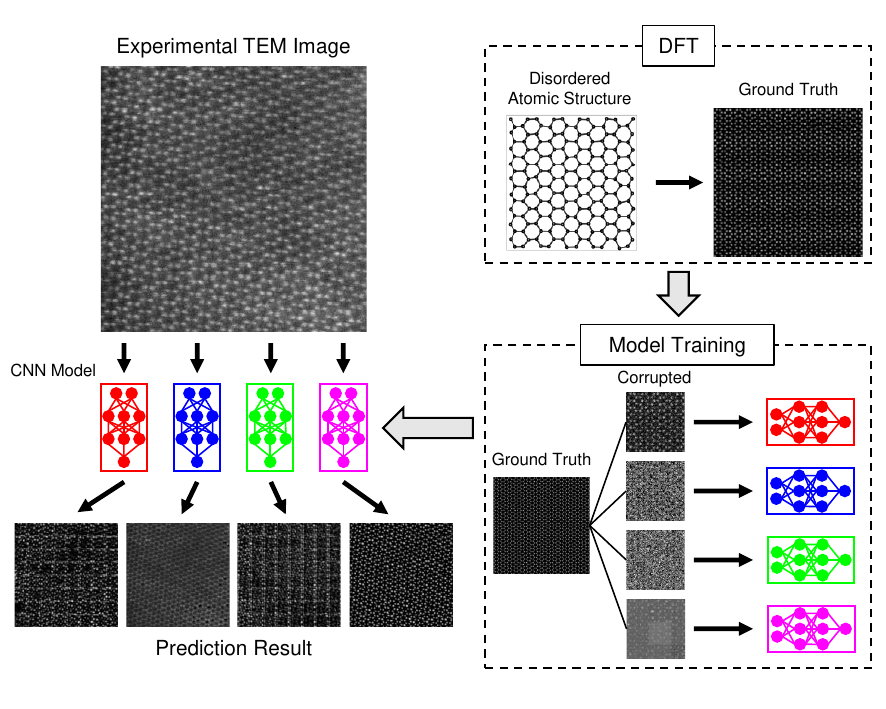}}
\caption{Schematic representation of our workflow.}
\label{fig:fig_S2}
\end{figure*}

Figure~\ref{fig:fig_S2} presents a schematic of our workflow. The ground truth (GT) image of the disordered atomic structure is a charge density map, which was obtained by DFT calculation. Each color of the CNN model represents that each model was trained by a different noise type. The CNN model in red (blue, green, and magenta) indicates that it uses G (SP, G\&SP, and BB\(+\)G) noise type for training.

\bibliography{manuscript}

\providecommand{\noopsort}[1]{}\providecommand{\singleletter}[1]{#1}%
\begin{thebibliography}{35}%
\makeatletter
\providecommand \@ifxundefined [1]{%
 \@ifx{#1\undefined}
}%
\providecommand \@ifnum [1]{%
 \ifnum #1\expandafter \@firstoftwo
 \else \expandafter \@secondoftwo
 \fi
}%
\providecommand \@ifx [1]{%
 \ifx #1\expandafter \@firstoftwo
 \else \expandafter \@secondoftwo
 \fi
}%
\providecommand \natexlab [1]{#1}%
\providecommand \enquote  [1]{``#1''}%
\providecommand \bibnamefont  [1]{#1}%
\providecommand \bibfnamefont [1]{#1}%
\providecommand \citenamefont [1]{#1}%
\providecommand \href@noop [0]{\@secondoftwo}%
\providecommand \href [0]{\begingroup \@sanitize@url \@href}%
\providecommand \@href[1]{\@@startlink{#1}\@@href}%
\providecommand \@@href[1]{\endgroup#1\@@endlink}%
\providecommand \@sanitize@url [0]{\catcode `\\12\catcode `\$12\catcode `\&12\catcode `\#12\catcode `\^12\catcode `\_12\catcode `\%12\relax}%
\providecommand \@@startlink[1]{}%
\providecommand \@@endlink[0]{}%
\providecommand \url  [0]{\begingroup\@sanitize@url \@url }%
\providecommand \@url [1]{\endgroup\@href {#1}{\urlprefix }}%
\providecommand \urlprefix  [0]{URL }%
\providecommand \Eprint [0]{\href }%
\providecommand \doibase [0]{https://doi.org/}%
\providecommand \selectlanguage [0]{\@gobble}%
\providecommand \bibinfo  [0]{\@secondoftwo}%
\providecommand \bibfield  [0]{\@secondoftwo}%
\providecommand \translation [1]{[#1]}%
\providecommand \BibitemOpen [0]{}%
\providecommand \bibitemStop [0]{}%
\providecommand \bibitemNoStop [0]{.\EOS\space}%
\providecommand \EOS [0]{\spacefactor3000\relax}%
\providecommand \BibitemShut  [1]{\csname bibitem#1\endcsname}%
\let\auto@bib@innerbib\@empty
\bibitem [{\citenamefont {Kloukiniotis}\ \emph {et~al.}(2022)\citenamefont {Kloukiniotis}, \citenamefont {Papandreou}, \citenamefont {Lalos}, \citenamefont {Kapsalas}, \citenamefont {Nguyen},\ and\ \citenamefont {Moustakas}}]{autonomous}%
  \BibitemOpen
  \bibfield  {author} {\bibinfo {author} {\bibfnamefont {A.}~\bibnamefont {Kloukiniotis}}, \bibinfo {author} {\bibfnamefont {A.}~\bibnamefont {Papandreou}}, \bibinfo {author} {\bibfnamefont {A.}~\bibnamefont {Lalos}}, \bibinfo {author} {\bibfnamefont {P.}~\bibnamefont {Kapsalas}}, \bibinfo {author} {\bibfnamefont {D.~V.}\ \bibnamefont {Nguyen}},\ and\ \bibinfo {author} {\bibfnamefont {K.}~\bibnamefont {Moustakas}},\ }\href {https://doi.org/10.1109/OJITS.2022.3142612} {\bibfield  {journal} {\bibinfo  {journal} {IEEE Open Journal of Intelligent Transportation Systems}\ }\textbf {\bibinfo {volume} {3}},\ \bibinfo {pages} {61} (\bibinfo {year} {2022})}\BibitemShut {NoStop}%
\bibitem [{\citenamefont {Moseley}\ \emph {et~al.}(2021)\citenamefont {Moseley}, \citenamefont {Bickel}, \citenamefont {Lopez-Francos},\ and\ \citenamefont {Rana}}]{Moseley_2021_CVPR}%
  \BibitemOpen
  \bibfield  {author} {\bibinfo {author} {\bibfnamefont {B.}~\bibnamefont {Moseley}}, \bibinfo {author} {\bibfnamefont {V.}~\bibnamefont {Bickel}}, \bibinfo {author} {\bibfnamefont {I.~G.}\ \bibnamefont {Lopez-Francos}},\ and\ \bibinfo {author} {\bibfnamefont {L.}~\bibnamefont {Rana}},\ }in\ \href@noop {} {\emph {\bibinfo {booktitle} {Proceedings of the IEEE/CVF Conference on Computer Vision and Pattern Recognition (CVPR)}}}\ (\bibinfo {year} {2021})\ pp.\ \bibinfo {pages} {6317--6327}\BibitemShut {NoStop}%
\bibitem [{\citenamefont {Karwowska}\ and\ \citenamefont {Wierzbicki}(2022)}]{satellite_review}%
  \BibitemOpen
  \bibfield  {author} {\bibinfo {author} {\bibfnamefont {K.}~\bibnamefont {Karwowska}}\ and\ \bibinfo {author} {\bibfnamefont {D.}~\bibnamefont {Wierzbicki}},\ }\href {https://doi.org/10.1109/JSTARS.2022.3167646} {\bibfield  {journal} {\bibinfo  {journal} {IEEE Journal of Selected Topics in Applied Earth Observations and Remote Sensing}\ }\textbf {\bibinfo {volume} {15}},\ \bibinfo {pages} {3292} (\bibinfo {year} {2022})}\BibitemShut {NoStop}%
\bibitem [{\citenamefont {Song}\ \emph {et~al.}(2021)\citenamefont {Song}, \citenamefont {Jeong}, \citenamefont {Park}, \citenamefont {Kim}, \citenamefont {Seo},\ and\ \citenamefont {Ye}}]{satellite_denoise}%
  \BibitemOpen
  \bibfield  {author} {\bibinfo {author} {\bibfnamefont {J.}~\bibnamefont {Song}}, \bibinfo {author} {\bibfnamefont {J.~H.}\ \bibnamefont {Jeong}}, \bibinfo {author} {\bibfnamefont {D.~S.}\ \bibnamefont {Park}}, \bibinfo {author} {\bibfnamefont {H.~H.}\ \bibnamefont {Kim}}, \bibinfo {author} {\bibfnamefont {D.~C.}\ \bibnamefont {Seo}},\ and\ \bibinfo {author} {\bibfnamefont {J.~C.}\ \bibnamefont {Ye}},\ }\href {https://doi.org/10.1109/TGRS.2020.3025601} {\bibfield  {journal} {\bibinfo  {journal} {IEEE Transactions on Geoscience and Remote Sensing}\ }\textbf {\bibinfo {volume} {59}},\ \bibinfo {pages} {6823} (\bibinfo {year} {2021})}\BibitemShut {NoStop}%
\bibitem [{\citenamefont {Coupe}\ \emph {et~al.}(2008)\citenamefont {Coupe}, \citenamefont {Yger}, \citenamefont {Prima}, \citenamefont {Hellier}, \citenamefont {Kervrann},\ and\ \citenamefont {Barillot}}]{Mean_filter}%
  \BibitemOpen
  \bibfield  {author} {\bibinfo {author} {\bibfnamefont {P.}~\bibnamefont {Coupe}}, \bibinfo {author} {\bibfnamefont {P.}~\bibnamefont {Yger}}, \bibinfo {author} {\bibfnamefont {S.}~\bibnamefont {Prima}}, \bibinfo {author} {\bibfnamefont {P.}~\bibnamefont {Hellier}}, \bibinfo {author} {\bibfnamefont {C.}~\bibnamefont {Kervrann}},\ and\ \bibinfo {author} {\bibfnamefont {C.}~\bibnamefont {Barillot}},\ }\href {https://doi.org/10.1109/TMI.2007.906087} {\bibfield  {journal} {\bibinfo  {journal} {IEEE Transactions on Medical Imaging}\ }\textbf {\bibinfo {volume} {27}},\ \bibinfo {pages} {425} (\bibinfo {year} {2008})}\BibitemShut {NoStop}%
\bibitem [{\citenamefont {Young}\ and\ \citenamefont {{van Vliet}}(1995)}]{Gaussian_filter}%
  \BibitemOpen
  \bibfield  {author} {\bibinfo {author} {\bibfnamefont {I.~T.}\ \bibnamefont {Young}}\ and\ \bibinfo {author} {\bibfnamefont {L.~J.}\ \bibnamefont {{van Vliet}}},\ }\href {https://doi.org/https://doi.org/10.1016/0165-1684(95)00020-E} {\bibfield  {journal} {\bibinfo  {journal} {Signal Processing}\ }\textbf {\bibinfo {volume} {44}},\ \bibinfo {pages} {139} (\bibinfo {year} {1995})}\BibitemShut {NoStop}%
\bibitem [{\citenamefont {Astola}\ \emph {et~al.}(1990)\citenamefont {Astola}, \citenamefont {Haavisto},\ and\ \citenamefont {Neuvo}}]{Median_filter}%
  \BibitemOpen
  \bibfield  {author} {\bibinfo {author} {\bibfnamefont {J.}~\bibnamefont {Astola}}, \bibinfo {author} {\bibfnamefont {P.}~\bibnamefont {Haavisto}},\ and\ \bibinfo {author} {\bibfnamefont {Y.}~\bibnamefont {Neuvo}},\ }\href {https://doi.org/10.1109/5.54807} {\bibfield  {journal} {\bibinfo  {journal} {Proceedings of the IEEE}\ }\textbf {\bibinfo {volume} {78}},\ \bibinfo {pages} {678} (\bibinfo {year} {1990})}\BibitemShut {NoStop}%
\bibitem [{\citenamefont {Baker}\ and\ \citenamefont {Kanade}(2002)}]{IEEE.Tran.Img.Proc.24.9.1167(2002)}%
  \BibitemOpen
  \bibfield  {author} {\bibinfo {author} {\bibfnamefont {S.}~\bibnamefont {Baker}}\ and\ \bibinfo {author} {\bibfnamefont {T.}~\bibnamefont {Kanade}},\ }\href {https://doi.org/10.1109/TPAMI.2002.1033210} {\bibfield  {journal} {\bibinfo  {journal} {IEEE Transactions on Pattern Analysis and Machine Intelligence}\ }\textbf {\bibinfo {volume} {24}},\ \bibinfo {pages} {1167} (\bibinfo {year} {2002})}\BibitemShut {NoStop}%
\bibitem [{\citenamefont {Lin}\ and\ \citenamefont {Shum}(2004)}]{IEEE.Tran.Patt.Anal.26.1.83(2004)}%
  \BibitemOpen
  \bibfield  {author} {\bibinfo {author} {\bibfnamefont {Z.}~\bibnamefont {Lin}}\ and\ \bibinfo {author} {\bibfnamefont {H.-Y.}\ \bibnamefont {Shum}},\ }\href {https://doi.org/10.1109/TPAMI.2004.1261081} {\bibfield  {journal} {\bibinfo  {journal} {IEEE Transactions on Pattern Analysis and Machine Intelligence}\ }\textbf {\bibinfo {volume} {26}},\ \bibinfo {pages} {83} (\bibinfo {year} {2004})}\BibitemShut {NoStop}%
\bibitem [{\citenamefont {Levin}\ and\ \citenamefont {Nadler}(2011)}]{CVPR.2833(2011)}%
  \BibitemOpen
  \bibfield  {author} {\bibinfo {author} {\bibfnamefont {A.}~\bibnamefont {Levin}}\ and\ \bibinfo {author} {\bibfnamefont {B.}~\bibnamefont {Nadler}},\ }in\ \href {https://doi.org/10.1109/CVPR.2011.5995309} {\emph {\bibinfo {booktitle} {CVPR 2011}}}\ (\bibinfo {year} {2011})\ pp.\ \bibinfo {pages} {2833--2840}\BibitemShut {NoStop}%
\bibitem [{\citenamefont {Knaus}\ and\ \citenamefont {Zwicker}(2014)}]{IEEE.Tran.Img.Proc.23.7.3114(2014)}%
  \BibitemOpen
  \bibfield  {author} {\bibinfo {author} {\bibfnamefont {C.}~\bibnamefont {Knaus}}\ and\ \bibinfo {author} {\bibfnamefont {M.}~\bibnamefont {Zwicker}},\ }\href {https://doi.org/10.1109/TIP.2014.2326771} {\bibfield  {journal} {\bibinfo  {journal} {IEEE Transactions on Image Processing}\ }\textbf {\bibinfo {volume} {23}},\ \bibinfo {pages} {3114} (\bibinfo {year} {2014})}\BibitemShut {NoStop}%
\bibitem [{\citenamefont {Fan}\ \emph {et~al.}(2019)\citenamefont {Fan}, \citenamefont {Zhang}, \citenamefont {Fan},\ and\ \citenamefont {Zhang}}]{Fan2019}%
  \BibitemOpen
  \bibfield  {author} {\bibinfo {author} {\bibfnamefont {L.}~\bibnamefont {Fan}}, \bibinfo {author} {\bibfnamefont {F.}~\bibnamefont {Zhang}}, \bibinfo {author} {\bibfnamefont {H.}~\bibnamefont {Fan}},\ and\ \bibinfo {author} {\bibfnamefont {C.}~\bibnamefont {Zhang}},\ }\href {https://doi.org/10.1186/s42492-019-0016-7} {\bibfield  {journal} {\bibinfo  {journal} {Visual Computing for Industry, Biomedicine, and Art}\ }\textbf {\bibinfo {volume} {2}},\ \bibinfo {pages} {7} (\bibinfo {year} {2019})}\BibitemShut {NoStop}%
\bibitem [{\citenamefont {Tian}\ \emph {et~al.}(2020)\citenamefont {Tian}, \citenamefont {Fei}, \citenamefont {Zheng}, \citenamefont {Xu}, \citenamefont {Zuo},\ and\ \citenamefont {Lin}}]{N.N.131.251(2020)}%
  \BibitemOpen
  \bibfield  {author} {\bibinfo {author} {\bibfnamefont {C.}~\bibnamefont {Tian}}, \bibinfo {author} {\bibfnamefont {L.}~\bibnamefont {Fei}}, \bibinfo {author} {\bibfnamefont {W.}~\bibnamefont {Zheng}}, \bibinfo {author} {\bibfnamefont {Y.}~\bibnamefont {Xu}}, \bibinfo {author} {\bibfnamefont {W.}~\bibnamefont {Zuo}},\ and\ \bibinfo {author} {\bibfnamefont {C.-W.}\ \bibnamefont {Lin}},\ }\href {https://doi.org/https://doi.org/10.1016/j.neunet.2020.07.025} {\bibfield  {journal} {\bibinfo  {journal} {Neural Networks}\ }\textbf {\bibinfo {volume} {131}},\ \bibinfo {pages} {251} (\bibinfo {year} {2020})}\BibitemShut {NoStop}%
\bibitem [{\citenamefont {Thakur}\ \emph {et~al.}(2021)\citenamefont {Thakur}, \citenamefont {Chatterjee}, \citenamefont {Yadav},\ and\ \citenamefont {Gupta}}]{IEEE.Acc.9(2021)}%
  \BibitemOpen
  \bibfield  {author} {\bibinfo {author} {\bibfnamefont {R.~S.}\ \bibnamefont {Thakur}}, \bibinfo {author} {\bibfnamefont {S.}~\bibnamefont {Chatterjee}}, \bibinfo {author} {\bibfnamefont {R.~N.}\ \bibnamefont {Yadav}},\ and\ \bibinfo {author} {\bibfnamefont {L.}~\bibnamefont {Gupta}},\ }\href {https://doi.org/10.1109/ACCESS.2021.3092425} {\bibfield  {journal} {\bibinfo  {journal} {IEEE Access}\ }\textbf {\bibinfo {volume} {9}},\ \bibinfo {pages} {93338} (\bibinfo {year} {2021})}\BibitemShut {NoStop}%
\bibitem [{\citenamefont {Zhang}\ \emph {et~al.}(2017)\citenamefont {Zhang}, \citenamefont {Zuo}, \citenamefont {Chen}, \citenamefont {Meng},\ and\ \citenamefont {Zhang}}]{IEEE.Tran.Img.Proc.26.3142(2017)}%
  \BibitemOpen
  \bibfield  {author} {\bibinfo {author} {\bibfnamefont {K.}~\bibnamefont {Zhang}}, \bibinfo {author} {\bibfnamefont {W.}~\bibnamefont {Zuo}}, \bibinfo {author} {\bibfnamefont {Y.}~\bibnamefont {Chen}}, \bibinfo {author} {\bibfnamefont {D.}~\bibnamefont {Meng}},\ and\ \bibinfo {author} {\bibfnamefont {L.}~\bibnamefont {Zhang}},\ }\href {https://doi.org/10.1109/TIP.2017.2662206} {\bibfield  {journal} {\bibinfo  {journal} {IEEE Transactions on Image Processing}\ }\textbf {\bibinfo {volume} {26}},\ \bibinfo {pages} {3142} (\bibinfo {year} {2017})}\BibitemShut {NoStop}%
\bibitem [{\citenamefont {Ho}\ \emph {et~al.}(2021)\citenamefont {Ho}, \citenamefont {Zhou},\ and\ \citenamefont {He}}]{dnCNN01}%
  \BibitemOpen
  \bibfield  {author} {\bibinfo {author} {\bibfnamefont {M.~M.}\ \bibnamefont {Ho}}, \bibinfo {author} {\bibfnamefont {J.}~\bibnamefont {Zhou}},\ and\ \bibinfo {author} {\bibfnamefont {G.}~\bibnamefont {He}},\ }\href {https://doi.org/10.1109/TIP.2020.3046872} {\bibfield  {journal} {\bibinfo  {journal} {IEEE Transactions on Image Processing}\ }\textbf {\bibinfo {volume} {30}},\ \bibinfo {pages} {1702} (\bibinfo {year} {2021})}\BibitemShut {NoStop}%
\bibitem [{\citenamefont {Murali}\ and\ \citenamefont {Sudeep}(2020)}]{dnCNN02}%
  \BibitemOpen
  \bibfield  {author} {\bibinfo {author} {\bibfnamefont {V.}~\bibnamefont {Murali}}\ and\ \bibinfo {author} {\bibfnamefont {P.~V.}\ \bibnamefont {Sudeep}},\ }in\ \href@noop {} {\emph {\bibinfo {booktitle} {Advances in Communication Systems and Networks}}}\ (\bibinfo  {publisher} {Springer Singapore},\ \bibinfo {address} {Singapore},\ \bibinfo {year} {2020})\ pp.\ \bibinfo {pages} {847--859}\BibitemShut {NoStop}%
\bibitem [{\citenamefont {Zhang}\ \emph {et~al.}(2018)\citenamefont {Zhang}, \citenamefont {Zuo},\ and\ \citenamefont {Zhang}}]{FFDNet}%
  \BibitemOpen
  \bibfield  {author} {\bibinfo {author} {\bibfnamefont {K.}~\bibnamefont {Zhang}}, \bibinfo {author} {\bibfnamefont {W.}~\bibnamefont {Zuo}},\ and\ \bibinfo {author} {\bibfnamefont {L.}~\bibnamefont {Zhang}},\ }\href {https://doi.org/10.1109/TIP.2018.2839891} {\bibfield  {journal} {\bibinfo  {journal} {IEEE Transactions on Image Processing}\ }\textbf {\bibinfo {volume} {27}},\ \bibinfo {pages} {4608} (\bibinfo {year} {2018})}\BibitemShut {NoStop}%
\bibitem [{\citenamefont {Lin}\ \emph {et~al.}(2021)\citenamefont {Lin}, \citenamefont {Zhang}, \citenamefont {Wang}, \citenamefont {Yang},\ and\ \citenamefont {Xin}}]{Sci.Rep.11.5386(2021)}%
  \BibitemOpen
  \bibfield  {author} {\bibinfo {author} {\bibfnamefont {R.}~\bibnamefont {Lin}}, \bibinfo {author} {\bibfnamefont {R.}~\bibnamefont {Zhang}}, \bibinfo {author} {\bibfnamefont {C.}~\bibnamefont {Wang}}, \bibinfo {author} {\bibfnamefont {X.-Q.}\ \bibnamefont {Yang}},\ and\ \bibinfo {author} {\bibfnamefont {H.~L.}\ \bibnamefont {Xin}},\ }\href {https://doi.org/10.1038/s41598-021-84499-w} {\bibfield  {journal} {\bibinfo  {journal} {Scientific Reports}\ }\textbf {\bibinfo {volume} {11}},\ \bibinfo {pages} {5386} (\bibinfo {year} {2021})}\BibitemShut {NoStop}%
\bibitem [{\citenamefont {Mohan}\ \emph {et~al.}(2022)\citenamefont {Mohan}, \citenamefont {Manzorro}, \citenamefont {Vincent}, \citenamefont {Tang}, \citenamefont {Sheth}, \citenamefont {Simoncelli}, \citenamefont {Matteson}, \citenamefont {Crozier},\ and\ \citenamefont {Fernandez-Granda}}]{IEEE.Tran.Comp.Img.8.585(2022)}%
  \BibitemOpen
  \bibfield  {author} {\bibinfo {author} {\bibfnamefont {S.}~\bibnamefont {Mohan}}, \bibinfo {author} {\bibfnamefont {R.}~\bibnamefont {Manzorro}}, \bibinfo {author} {\bibfnamefont {J.~L.}\ \bibnamefont {Vincent}}, \bibinfo {author} {\bibfnamefont {B.}~\bibnamefont {Tang}}, \bibinfo {author} {\bibfnamefont {D.~Y.}\ \bibnamefont {Sheth}}, \bibinfo {author} {\bibfnamefont {E.~P.}\ \bibnamefont {Simoncelli}}, \bibinfo {author} {\bibfnamefont {D.~S.}\ \bibnamefont {Matteson}}, \bibinfo {author} {\bibfnamefont {P.~A.}\ \bibnamefont {Crozier}},\ and\ \bibinfo {author} {\bibfnamefont {C.}~\bibnamefont {Fernandez-Granda}},\ }\href {https://doi.org/10.1109/TCI.2022.3176536} {\bibfield  {journal} {\bibinfo  {journal} {IEEE Transactions on Computational Imaging}\ }\textbf {\bibinfo {volume} {8}},\ \bibinfo {pages} {585} (\bibinfo {year} {2022})}\BibitemShut {NoStop}%
\bibitem [{\citenamefont {Joucken}\ \emph {et~al.}(2022)\citenamefont {Joucken}, \citenamefont {Davenport}, \citenamefont {Ge}, \citenamefont {Quezada-Lopez}, \citenamefont {Taniguchi}, \citenamefont {Watanabe}, \citenamefont {Velasco}, \citenamefont {Lagoute},\ and\ \citenamefont {Kaindl}}]{Phys.Rev.Mat.6.123802(2022)}%
  \BibitemOpen
  \bibfield  {author} {\bibinfo {author} {\bibfnamefont {F.}~\bibnamefont {Joucken}}, \bibinfo {author} {\bibfnamefont {J.~L.}\ \bibnamefont {Davenport}}, \bibinfo {author} {\bibfnamefont {Z.}~\bibnamefont {Ge}}, \bibinfo {author} {\bibfnamefont {E.~A.}\ \bibnamefont {Quezada-Lopez}}, \bibinfo {author} {\bibfnamefont {T.}~\bibnamefont {Taniguchi}}, \bibinfo {author} {\bibfnamefont {K.}~\bibnamefont {Watanabe}}, \bibinfo {author} {\bibfnamefont {J.}~\bibnamefont {Velasco}}, \bibinfo {author} {\bibfnamefont {J.}~\bibnamefont {Lagoute}},\ and\ \bibinfo {author} {\bibfnamefont {R.~A.}\ \bibnamefont {Kaindl}},\ }\href {https://doi.org/10.1103/PhysRevMaterials.6.123802} {\bibfield  {journal} {\bibinfo  {journal} {Phys. Rev. Mater.}\ }\textbf {\bibinfo {volume} {6}},\ \bibinfo {pages} {123802} (\bibinfo {year} {2022})}\BibitemShut {NoStop}%
\bibitem [{\citenamefont {Fan}\ \emph {et~al.}(2023)\citenamefont {Fan}, \citenamefont {Pan}, \citenamefont {Huang}, \citenamefont {Xu},\ and\ \citenamefont {Wang}}]{ICRAIC(2023)290}%
  \BibitemOpen
  \bibfield  {author} {\bibinfo {author} {\bibfnamefont {X.}~\bibnamefont {Fan}}, \bibinfo {author} {\bibfnamefont {H.}~\bibnamefont {Pan}}, \bibinfo {author} {\bibfnamefont {Y.}~\bibnamefont {Huang}}, \bibinfo {author} {\bibfnamefont {Y.}~\bibnamefont {Xu}},\ and\ \bibinfo {author} {\bibfnamefont {X.}~\bibnamefont {Wang}},\ }in\ \href {https://doi.org/10.1109/ICRAIC61978.2023.00059} {\emph {\bibinfo {booktitle} {2023 3rd International Conference on Robotics, Automation and Intelligent Control (ICRAIC)}}}\ (\bibinfo {year} {2023})\ pp.\ \bibinfo {pages} {290--294}\BibitemShut {NoStop}%
\bibitem [{\citenamefont {Zhu}\ \emph {et~al.}(2024)\citenamefont {Zhu}, \citenamefont {Lu}, \citenamefont {Yuan}, \citenamefont {He}, \citenamefont {Zheng}, \citenamefont {Jiang}, \citenamefont {Yan},\ and\ \citenamefont {Sun}}]{JPCL.15.7.1985(2024)}%
  \BibitemOpen
  \bibfield  {author} {\bibinfo {author} {\bibfnamefont {Z.}~\bibnamefont {Zhu}}, \bibinfo {author} {\bibfnamefont {J.}~\bibnamefont {Lu}}, \bibinfo {author} {\bibfnamefont {S.}~\bibnamefont {Yuan}}, \bibinfo {author} {\bibfnamefont {Y.}~\bibnamefont {He}}, \bibinfo {author} {\bibfnamefont {F.}~\bibnamefont {Zheng}}, \bibinfo {author} {\bibfnamefont {H.}~\bibnamefont {Jiang}}, \bibinfo {author} {\bibfnamefont {Y.}~\bibnamefont {Yan}},\ and\ \bibinfo {author} {\bibfnamefont {Q.}~\bibnamefont {Sun}},\ }\href {https://doi.org/10.1021/acs.jpclett.3c03504} {\bibfield  {journal} {\bibinfo  {journal} {The Journal of Physical Chemistry Letters}\ }\textbf {\bibinfo {volume} {15}},\ \bibinfo {pages} {1985} (\bibinfo {year} {2024})}\BibitemShut {NoStop}%
\bibitem [{\citenamefont {Zhang}\ \emph {et~al.}(2020)\citenamefont {Zhang}, \citenamefont {Han}, \citenamefont {Zhang},\ and\ \citenamefont {Voyles}}]{Ultra.219.113123(2020)}%
  \BibitemOpen
  \bibfield  {author} {\bibinfo {author} {\bibfnamefont {C.}~\bibnamefont {Zhang}}, \bibinfo {author} {\bibfnamefont {R.}~\bibnamefont {Han}}, \bibinfo {author} {\bibfnamefont {A.~R.}\ \bibnamefont {Zhang}},\ and\ \bibinfo {author} {\bibfnamefont {P.~M.}\ \bibnamefont {Voyles}},\ }\href {https://doi.org/https://doi.org/10.1016/j.ultramic.2020.113123} {\bibfield  {journal} {\bibinfo  {journal} {Ultramicroscopy}\ }\textbf {\bibinfo {volume} {219}},\ \bibinfo {pages} {113123} (\bibinfo {year} {2020})}\BibitemShut {NoStop}%
\bibitem [{\citenamefont {Wang}\ \emph {et~al.}(2020)\citenamefont {Wang}, \citenamefont {Henninen}, \citenamefont {Keller},\ and\ \citenamefont {Erni}}]{Appl.Mic.50.23(2020)}%
  \BibitemOpen
  \bibfield  {author} {\bibinfo {author} {\bibfnamefont {F.}~\bibnamefont {Wang}}, \bibinfo {author} {\bibfnamefont {T.~R.}\ \bibnamefont {Henninen}}, \bibinfo {author} {\bibfnamefont {D.}~\bibnamefont {Keller}},\ and\ \bibinfo {author} {\bibfnamefont {R.}~\bibnamefont {Erni}},\ }\href {https://doi.org/10.1186/s42649-020-00041-8} {\bibfield  {journal} {\bibinfo  {journal} {Applied Microscopy}\ }\textbf {\bibinfo {volume} {50}},\ \bibinfo {pages} {23} (\bibinfo {year} {2020})}\BibitemShut {NoStop}%
\bibitem [{\citenamefont {Ozaki}(2003)}]{Phys.Rev.B.67.155108(2003)}%
  \BibitemOpen
  \bibfield  {author} {\bibinfo {author} {\bibfnamefont {T.}~\bibnamefont {Ozaki}},\ }\href {https://doi.org/10.1103/PhysRevB.67.155108} {\bibfield  {journal} {\bibinfo  {journal} {Phys. Rev. B}\ }\textbf {\bibinfo {volume} {67}},\ \bibinfo {pages} {155108} (\bibinfo {year} {2003})}\BibitemShut {NoStop}%
\bibitem [{\citenamefont {Hamann}\ \emph {et~al.}(1979)\citenamefont {Hamann}, \citenamefont {Schl\"uter},\ and\ \citenamefont {Chiang}}]{Phys.Rev.Lett.43.1494(1979)}%
  \BibitemOpen
  \bibfield  {author} {\bibinfo {author} {\bibfnamefont {D.~R.}\ \bibnamefont {Hamann}}, \bibinfo {author} {\bibfnamefont {M.}~\bibnamefont {Schl\"uter}},\ and\ \bibinfo {author} {\bibfnamefont {C.}~\bibnamefont {Chiang}},\ }\href {https://doi.org/10.1103/PhysRevLett.43.1494} {\bibfield  {journal} {\bibinfo  {journal} {Phys. Rev. Lett.}\ }\textbf {\bibinfo {volume} {43}},\ \bibinfo {pages} {1494} (\bibinfo {year} {1979})}\BibitemShut {NoStop}%
\bibitem [{\citenamefont {Kleinman}\ and\ \citenamefont {Bylander}(1982)}]{Phys.Rev.Lett.48.1425(1982)}%
  \BibitemOpen
  \bibfield  {author} {\bibinfo {author} {\bibfnamefont {L.}~\bibnamefont {Kleinman}}\ and\ \bibinfo {author} {\bibfnamefont {D.~M.}\ \bibnamefont {Bylander}},\ }\href {https://doi.org/10.1103/PhysRevLett.48.1425} {\bibfield  {journal} {\bibinfo  {journal} {Phys. Rev. Lett.}\ }\textbf {\bibinfo {volume} {48}},\ \bibinfo {pages} {1425} (\bibinfo {year} {1982})}\BibitemShut {NoStop}%
\bibitem [{\citenamefont {Perdew}\ \emph {et~al.}(1996)\citenamefont {Perdew}, \citenamefont {Burke},\ and\ \citenamefont {Ernzerhof}}]{Phys.Rev.Lett.77.3865(1996)}%
  \BibitemOpen
  \bibfield  {author} {\bibinfo {author} {\bibfnamefont {J.~P.}\ \bibnamefont {Perdew}}, \bibinfo {author} {\bibfnamefont {K.}~\bibnamefont {Burke}},\ and\ \bibinfo {author} {\bibfnamefont {M.}~\bibnamefont {Ernzerhof}},\ }\href {https://doi.org/10.1103/PhysRevLett.77.3865} {\bibfield  {journal} {\bibinfo  {journal} {Phys. Rev. Lett.}\ }\textbf {\bibinfo {volume} {77}},\ \bibinfo {pages} {3865} (\bibinfo {year} {1996})}\BibitemShut {NoStop}%
\bibitem [{\citenamefont {Chollet}\ \emph {et~al.}(2015)\citenamefont {Chollet} \emph {et~al.}}]{KERAS}%
  \BibitemOpen
  \bibfield  {author} {\bibinfo {author} {\bibfnamefont {F.}~\bibnamefont {Chollet}} \emph {et~al.},\ }\href@noop {} {\bibinfo {title} {Keras}},\ \bibinfo {howpublished} {\url{https://keras.io}} (\bibinfo {year} {2015})\BibitemShut {NoStop}%
\bibitem [{\citenamefont {Abadi}\ \emph {et~al.}(2015)\citenamefont {Abadi}, \citenamefont {Agarwal}, \citenamefont {Barham}, \citenamefont {Brevdo}, \citenamefont {Chen}, \citenamefont {Citro}, \citenamefont {Corrado}, \citenamefont {Davis}, \citenamefont {Dean}, \citenamefont {Devin}, \citenamefont {Ghemawat}, \citenamefont {Goodfellow}, \citenamefont {Harp}, \citenamefont {Irving}, \citenamefont {Isard}, \citenamefont {Jia}, \citenamefont {Jozefowicz}, \citenamefont {Kaiser}, \citenamefont {Kudlur}, \citenamefont {Levenberg}, \citenamefont {Man\'{e}}, \citenamefont {Monga}, \citenamefont {Moore}, \citenamefont {Murray}, \citenamefont {Olah}, \citenamefont {Schuster}, \citenamefont {Shlens}, \citenamefont {Steiner}, \citenamefont {Sutskever}, \citenamefont {Talwar}, \citenamefont {Tucker}, \citenamefont {Vanhoucke}, \citenamefont {Vasudevan}, \citenamefont {Vi\'{e}gas}, \citenamefont {Vinyals}, \citenamefont {Warden}, \citenamefont {Wattenberg}, \citenamefont {Wicke}, \citenamefont {Yu},\ and\ \citenamefont
  {Zheng}}]{tensorflow}%
  \BibitemOpen
  \bibfield  {author} {\bibinfo {author} {\bibfnamefont {M.}~\bibnamefont {Abadi}}, \bibinfo {author} {\bibfnamefont {A.}~\bibnamefont {Agarwal}}, \bibinfo {author} {\bibfnamefont {P.}~\bibnamefont {Barham}}, \bibinfo {author} {\bibfnamefont {E.}~\bibnamefont {Brevdo}}, \bibinfo {author} {\bibfnamefont {Z.}~\bibnamefont {Chen}}, \bibinfo {author} {\bibfnamefont {C.}~\bibnamefont {Citro}}, \bibinfo {author} {\bibfnamefont {G.~S.}\ \bibnamefont {Corrado}}, \bibinfo {author} {\bibfnamefont {A.}~\bibnamefont {Davis}}, \bibinfo {author} {\bibfnamefont {J.}~\bibnamefont {Dean}}, \bibinfo {author} {\bibfnamefont {M.}~\bibnamefont {Devin}}, \bibinfo {author} {\bibfnamefont {S.}~\bibnamefont {Ghemawat}}, \bibinfo {author} {\bibfnamefont {I.}~\bibnamefont {Goodfellow}}, \bibinfo {author} {\bibfnamefont {A.}~\bibnamefont {Harp}}, \bibinfo {author} {\bibfnamefont {G.}~\bibnamefont {Irving}}, \bibinfo {author} {\bibfnamefont {M.}~\bibnamefont {Isard}}, \bibinfo {author} {\bibfnamefont {Y.}~\bibnamefont {Jia}}, \bibinfo
  {author} {\bibfnamefont {R.}~\bibnamefont {Jozefowicz}}, \bibinfo {author} {\bibfnamefont {L.}~\bibnamefont {Kaiser}}, \bibinfo {author} {\bibfnamefont {M.}~\bibnamefont {Kudlur}}, \bibinfo {author} {\bibfnamefont {J.}~\bibnamefont {Levenberg}}, \bibinfo {author} {\bibfnamefont {D.}~\bibnamefont {Man\'{e}}}, \bibinfo {author} {\bibfnamefont {R.}~\bibnamefont {Monga}}, \bibinfo {author} {\bibfnamefont {S.}~\bibnamefont {Moore}}, \bibinfo {author} {\bibfnamefont {D.}~\bibnamefont {Murray}}, \bibinfo {author} {\bibfnamefont {C.}~\bibnamefont {Olah}}, \bibinfo {author} {\bibfnamefont {M.}~\bibnamefont {Schuster}}, \bibinfo {author} {\bibfnamefont {J.}~\bibnamefont {Shlens}}, \bibinfo {author} {\bibfnamefont {B.}~\bibnamefont {Steiner}}, \bibinfo {author} {\bibfnamefont {I.}~\bibnamefont {Sutskever}}, \bibinfo {author} {\bibfnamefont {K.}~\bibnamefont {Talwar}}, \bibinfo {author} {\bibfnamefont {P.}~\bibnamefont {Tucker}}, \bibinfo {author} {\bibfnamefont {V.}~\bibnamefont {Vanhoucke}}, \bibinfo {author}
  {\bibfnamefont {V.}~\bibnamefont {Vasudevan}}, \bibinfo {author} {\bibfnamefont {F.}~\bibnamefont {Vi\'{e}gas}}, \bibinfo {author} {\bibfnamefont {O.}~\bibnamefont {Vinyals}}, \bibinfo {author} {\bibfnamefont {P.}~\bibnamefont {Warden}}, \bibinfo {author} {\bibfnamefont {M.}~\bibnamefont {Wattenberg}}, \bibinfo {author} {\bibfnamefont {M.}~\bibnamefont {Wicke}}, \bibinfo {author} {\bibfnamefont {Y.}~\bibnamefont {Yu}},\ and\ \bibinfo {author} {\bibfnamefont {X.}~\bibnamefont {Zheng}},\ }\href {https://www.tensorflow.org/} {\bibinfo {title} {{TensorFlow}: Large-scale machine learning on heterogeneous systems}} (\bibinfo {year} {2015}),\ \bibinfo {note} {software available from tensorflow.org}\BibitemShut {NoStop}%
\bibitem [{\citenamefont {Ronneberger}\ \emph {et~al.}(2015)\citenamefont {Ronneberger}, \citenamefont {Fischer},\ and\ \citenamefont {Brox}}]{U-Net}%
  \BibitemOpen
  \bibfield  {author} {\bibinfo {author} {\bibfnamefont {O.}~\bibnamefont {Ronneberger}}, \bibinfo {author} {\bibfnamefont {P.}~\bibnamefont {Fischer}},\ and\ \bibinfo {author} {\bibfnamefont {T.}~\bibnamefont {Brox}},\ }in\ \href@noop {} {\emph {\bibinfo {booktitle} {Medical Image Computing and Computer-Assisted Intervention -- MICCAI 2015}}}\ (\bibinfo  {publisher} {Springer International Publishing},\ \bibinfo {year} {2015})\ pp.\ \bibinfo {pages} {234--241}\BibitemShut {NoStop}%
\bibitem [{\citenamefont {Kingma}\ and\ \citenamefont {Ba}(2017)}]{ADAM}%
  \BibitemOpen
  \bibfield  {author} {\bibinfo {author} {\bibfnamefont {D.~P.}\ \bibnamefont {Kingma}}\ and\ \bibinfo {author} {\bibfnamefont {J.}~\bibnamefont {Ba}},\ }\href {https://arxiv.org/abs/1412.6980} {\bibinfo {title} {Adam: A method for stochastic optimization}} (\bibinfo {year} {2017}),\ \Eprint {https://arxiv.org/abs/1412.6980} {arXiv:1412.6980 [cs.LG]} \BibitemShut {NoStop}%
\bibitem [{\citenamefont {Wang}\ \emph {et~al.}(2003)\citenamefont {Wang}, \citenamefont {Simoncelli},\ and\ \citenamefont {Bovik}}]{SSIM}%
  \BibitemOpen
  \bibfield  {author} {\bibinfo {author} {\bibfnamefont {Z.}~\bibnamefont {Wang}}, \bibinfo {author} {\bibfnamefont {E.}~\bibnamefont {Simoncelli}},\ and\ \bibinfo {author} {\bibfnamefont {A.}~\bibnamefont {Bovik}},\ }in\ \href {https://doi.org/10.1109/ACSSC.2003.1292216} {\emph {\bibinfo {booktitle} {The Thrity-Seventh Asilomar Conference on Signals, Systems \& Computers, 2003}}},\ Vol.~\bibinfo {volume} {2}\ (\bibinfo {year} {2003})\ pp.\ \bibinfo {pages} {1398--1402}\BibitemShut {NoStop}%
\bibitem [{\citenamefont {Zhao}\ \emph {et~al.}(2017)\citenamefont {Zhao}, \citenamefont {Gallo}, \citenamefont {Frosio},\ and\ \citenamefont {Kautz}}]{IEEE.Trans.Comp.Imag.3.1.47}%
  \BibitemOpen
  \bibfield  {author} {\bibinfo {author} {\bibfnamefont {H.}~\bibnamefont {Zhao}}, \bibinfo {author} {\bibfnamefont {O.}~\bibnamefont {Gallo}}, \bibinfo {author} {\bibfnamefont {I.}~\bibnamefont {Frosio}},\ and\ \bibinfo {author} {\bibfnamefont {J.}~\bibnamefont {Kautz}},\ }\href {https://doi.org/10.1109/TCI.2016.2644865} {\bibfield  {journal} {\bibinfo  {journal} {IEEE Transactions on Computational Imaging}\ }\textbf {\bibinfo {volume} {3}},\ \bibinfo {pages} {47} (\bibinfo {year} {2017})}\BibitemShut {NoStop}%
\end{thebibliography}%

\end{document}